\documentclass[onecolumn,authoryear]{els-mrw}

\usepackage{amsmath,amssymb,amsfonts,amsthm,makeidx,graphicx}
\usepackage{txfonts}
\usepackage{helvet}

\begin{document}

\chapter{Brightest Cluster Galaxies and the Intracluster Light}\label{chapBCGICL}

\author[1]{Emanuele Contini}%
\author[1]{Sukyoung K. Yi}%
\author[1]{Seyoung Jeon}%

\address[1]{\orgname{Yonsei University}, \orgdiv{Department of Astronomy and Yonsei University Observatory}, \orgaddress{50 Yonsei-ro, Seodaemun-gu, Seoul 03722, Republic of Korea}}


\maketitle

Emanuele Contini: \email{emanuele.contini82@gmail.com}

\begin{glossary}[Glossary]
\term{Brightest cluster galaxy} the main galaxy in the center of a galaxy cluster.

\term{Intracluster Light} Diffuse light from stars not bound to any galaxy in within the cluster.

\term{Stellar stripping} Process acting on satellite galaxies due to tidal interactions with the host halo.
\end{glossary}
\begin{glossary}[Nomenclature]
\begin{tabular}{@{}lp{34pc}@{}}
BCG &Brightest Cluster Galaxy\\
BGG &Brightest Group Galaxy\\
ICL &Intra-Cluster Light\\
FOF &Friend-of-Friend\\
\end{tabular}
\end{glossary}

\begin{abstract}[Abstract]
In this chapter, we delve into the formation and primary characteristics of two significant components within galaxy clusters: the brightest cluster galaxies (BCGs) and the intracluster light (ICL). Drawing upon recent and pertinent studies in the field, we explore the mechanisms driving their growth from high redshift to the present day, i.e., mergers and stellar stripping. Mergers between satellite galaxies and the BCGs account for a significant amount of ICL, as well as stellar stripping which is responsible for the formation of the bulk of it. We also examine how these formation mechanisms are intertwined with the dynamical state of their host clusters, suggesting their potential utility as luminous tracers of dark matter.
\end{abstract}

\begin{BoxTypeA}[chapBCGICL:box1]{Key Points}

\begin{itemize}
\item BCGs: these galaxies undergo multiple phases of growth. In the early Universe, abundant cold gas facilitates their growth in mass through intense star formation. In a second step, due to hierarchical clustering, satellite galaxies of varying types, masses and morphology merge with BCGs. In later stages, this rapid growth via mergers transitions to what is termed as smooth accretion, involving the gradual incorporation of dwarf galaxies or significantly smaller objects.
\item Main formation channels of the ICL: mergers between satellites and the BCG, stellar stripping of stars from satellites due to tidal forces during their orbits, and the
accretion or pre-processing of material from external sources (formed through mergers and/or stellar stripping).
\item BCG+ICL can serve as luminous tracers of the dark matter distribution, particularly the ICL, and are linked to the dynamical state of their host.

\end{itemize}

\end{BoxTypeA}

\section{Introduction}\label{chapBCGICL:intro}
Galaxy clusters are the largest virialized structures in the Universe, extraordinary laboratories for studying galaxy formation and evolution. Although the amount of baryonic matter in the Universe is very low, around 5\%, clusters host galaxies of different mass, size, and morphology. Among these galaxies, the brightest cluster galaxies (BCGs) stand out as particularly peculiar.
Situated at the centers of the potential wells of dark matter halos, BCGs possess properties that significantly differ from those of other member galaxies, typically referred to as satellites.
Since the \emph{galactic cannibalism} theory (\citealt{ostriker-tremaine1975}) has been proposed, according to which central galaxies accrete satellites through mergers, it became evident that BCGs constitute the final stage of the evolution of galaxies that reside in the center of dark matter halos, where they grow in mass by acquiring the stellar and gaseous material coming from satellites that inevitably merge with them.

Stars in galaxies are bound to the potential well of the galaxy itself and, under some circumstances, they can be stripped and become part of what we call intracluster light (ICL). The ICL is an important component of galaxy clusters that is made of stars not bound to any member galaxy and that feel only the potential well of the dark matter halo (e.g., \citealt{contini2021,montes2022}). It was first discovered by \cite{zwicky1937}, but only during the past two decades the scientific community has spent a remarkable effort in studying the properties, and their implications (see, e.g., the reviews by \citealt{montes2022,arnaboldi2022}). The physical mechanisms responsible for the ICL formation are strictly linked to the evolution of the BCGs, i.e., from the satellite galaxies that orbit in the clusters and merge with the BCG. Meantime, tidal forces acting on orbiting satellites can be very effective, such that some of their mass can be lost in the process or they can even be totally disrupted. This stellar material is what the ICL is made of, unbound stars that leave the galaxy where they were born.

BCGs and ICL have a very common formation history, in the sense that their formation and assembly are strictly connected through physical processes such as the stellar material coming from satellite galaxies orbiting around the potential well of the cluster and that eventually merge with the BCG. Among this processes, which will be discussed in Section \ref{chapBCGICL:ICL_fm}, three of them are fundamental for both the growth of the BCG and the ICL, namely: mergers (see Def.\ref{def:mergers}, \citealt{monaco2006,murante2007,contini2014,contini2024a}), stellar stripping due to tidal interactions between satellites and the BCG (\citealt{rudick2009,rudick2011,contini2018,demaio2018,montes2018,contini2024b}), and accretion of stellar material from outside the virialized region, also called pre-processing (\citealt{mihos2005,sommer-larsen2006,ragusa2023,joo2023,contini2024b}). Given their intimate connection and the unclear separation between them (we will come back to this point in the following), these two components are often treated as a single system, and the only distinction is purely based either on the distance from the cluster center (e.g., \citealt{pillepich2018,contreras-santos2024}), or an isophote with a given surface brightness limit (e.g., \citealt{zibetti2005,montes2021}).

One approach to discerning the transition between BCG-bound stars and ICL is by fitting the combined BCG+ICL profile distribution with a multiple Sersic (\citealt{sersic1968}) profile (e.g., \citealt{kravtsov2018,zhang2019}). The innermost distribution represents the BCG, the outermost the ICL, and the transition region between the two can be viewed as the stellar halo typically surrounding the BCG, particularly evident in nearby clusters (\citealt{longobardi2015}). However, although BCG and ICL have common formation histories, they are quite different in terms of properties. Therefore, another approach to delineating these components involves examining their characteristics, such as colors, ages, and metallicity (\citealt{morishita2017,montes2018,demaio2018,contini2019}). For instance,
if the ICL primarily comprises stars from stripped satellite galaxies, its color, age, and metallicity would differ from those of the BCG. In such cases, the ICL would likely appear bluer, younger, and more metal-poor compared to the BCG. Conversely, in scenarios where mergers are the primary mechanism, gradients in colors (or age and metallicity) would not be expected (e.g., \citealt{contini2019}).

This chapter serves as an introduction to the primary mechanisms governing the formation and evolution of BCGs and ICL, along with their key properties. Our approach integrates observational evidence with theoretical predictions from simulations and analytic models, which have now achieved remarkable accuracy in matching the observed properties of galaxies in general. In Section \ref{chapBCGICL:BCGs}, we provide a brief overview of the BCGs formation, delving into the main theoretical aspects underlying their assembly. Section \ref{chapBCGICL:ICL} is dedicated to the formation of the ICL, where we analyze various mechanisms potentially responsible for its formation alongside observed properties such as colors, age, and metallicity. Given that many studies treat BCGs and ICL as a unified system, as will be elucidated in subsequent sections, Section \ref{chapBCGICL:BCGICL} focuses on the combined BCG+ICL, considering their properties and proposed procedures for reliably separating the two components. We conclude the chapter in Sections \ref{chapBCGICL:ICL_distr} and \ref{chapBCGICL:BCGICL_coev}, exploring the utility of BCG/ICL distributions in shedding light on dark matter and the evolution of these components in the latter stages of the evolution of the Universe. Finally, Section \ref{chapBCGICL:conclusions} provides a summary of the main points covered in the chapter.

\section{Brightest cluster galaxies}\label{chapBCGICL:BCGs}
BCGs are among the most massive objects in galaxy clusters, and among the most massive and luminous galaxies in the Universe. As mentioned in the introduction, these galaxies have formation and assembly histories that differ quite significantly from the rest of the galaxies that we know (\citealt{contini2014} and reference therein). The pioneering work by \cite{delucia2007} elucidated that BCGs at the cores of their dark matter halos form hierarchically, meaning they result from the merging of smaller building blocks provided by satellite galaxies orbiting within the potential well of the cluster. These galaxies undergo multiple phases of growth. In the early Universe, abundant cold gas facilitates their mass increase through intense star formation. Subsequently, due to hierarchical clustering, galaxies merge to form larger entities, with BCGs benefiting most by assimilating mass from merging satellite galaxies of varying types, masses, and morphologies. In later stages, this rapid growth via mergers transitions to what is termed as \emph{smooth accretion}, involving the gradual incorporation of dwarf galaxies or significantly smaller objects.

The duration of each growth phase and its dependence on external factors, such as the typical environment in which BCGs reside, remain unclear. It is plausible to assume that BCGs in smaller dark matter halos have had fewer opportunities to grow through mergers compared to those in larger halos, simply due to differences in the availability of building blocks between the two scenarios. Particularly in observational studies (\citealt{lidman2012,lin2013,oliva-altamirano2014}), the growth rate during the latter half of the Universe's lifespan is debated. While numerical simulations and semi-analytic models
(\citealt{murante2007,contini2018,ragone2018}) suggest similar growth rates, observations report varying figures, ranging from 100\% (\citealt{lidman2012}) to 50\% (\citealt{lin2013}), with instances of no growth observed as well (e.g., \citealt{oliva-altamirano2014}).

BCGs are also found in galaxy groups, where they are referred to as brightest group galaxies (BGGs). However, it is common to use the general term 'BCGs' without clear differentiation between galaxy groups and clusters. Nevertheless, BGGs and BCGs exhibit differences in their properties. Both simulations and observations have revealed that, on average, BGGs are less luminous, less massive, and have lower stellar velocity dispersions compared to BCGs (\citealt{sohn2020,marini2021,einasto2022}). The properties of BCGs are crucial in understanding their host clusters. Several studies (\citealt{shaaban2023} and references therein) have investigated the connection between BCG properties and those of their host clusters, examining surface brightness profiles, position angles, alignments, X-ray emissions, size, and others. Despite their names, it must be noted that BCGs are not always the central galaxies of their hosts (\citealt{kluge2020}) and, on the other way around, central galaxies are not always the brightest ones. It has been estimated that between 20\% and 40\% of central galaxies are not the brightest (\citealt{skibba2011,hoshino2015}), and M87 in Virgo is a prominent example.

An essential aspect of galaxy formation and evolution, applicable to all galaxies but particularly significant for BCGs, is the differentiation between the concepts of formation and assembly. \cite{delucia2007} provided clear insights into these concepts, demonstrating that the disparity between formation and assembly times is more pronounced for BCGs compared to other galaxies. The assembly time is defined as the time when the main progenitor (i.e., the main branch of the merger tree) has half of the final mass of the BCG at $z=0$, while the formation time includes in the definition all the progenitors of the BCG. \cite{delucia2007} showed that, while BCGs can form pretty early ($z\sim 5$), half of the stellar mass is actually assembled much later ($z\sim 0.5$).

\begin{figure}[t]
\centering
\includegraphics[angle=180,origin=c,width=.9\textwidth]{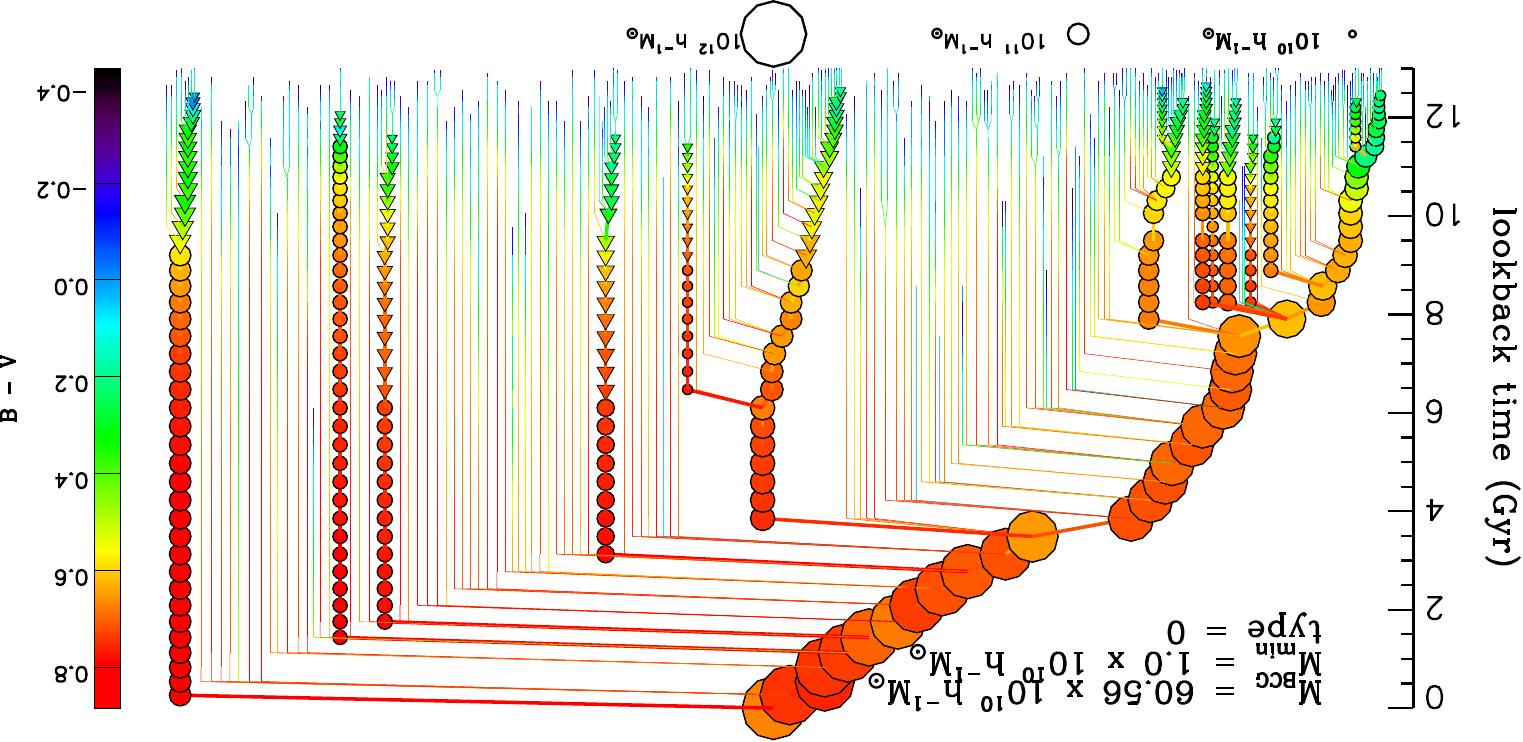}
\caption{An example of a BCG merger tree is illustrated in the figure. Circles represent galaxies that currently reside within the FOF group of the main progenitor of the BCG, while triangles represent galaxies that have not yet joined this group. The symbols are color-coded based on the B-V color, and their size scales with the stellar mass of the galaxies. The main progenitor branch, referred to in the text, is depicted as the left-most branch in the figure. Credit: \cite{delucia2007}.}
\label{chapBCGICL:tree}
\end{figure}

A nice example of a BCG merger tree is illustrated in Figure \ref{chapBCGICL:tree}, where circles refer to galaxies residing in the FOF group of the main branch, and triangles those that have not yet joined it. Symbols are also color coded based on the B-V color, an indicator of age and metallicity, and their size scales with the stellar mass. The main progenitor branch is the left-most one. This visualization effectively demonstrates the discrepancy between the two defined times, especially noticeable for BCGs, depending on whether all branches or only the main one are considered. It showcases that a significant portion of stars forms early, within separate objects, before eventually being assembled into the final BCG. This study remains a milestone in the field of hierarchical clustering, as it underscores the necessity of a hierarchical framework to fully comprehend the history of BCGs.

The concepts of formation and assembly are crucial in understanding the formation and evolution of the ICL as well. Just as the assembly of BCGs is intimately linked to that of their host halos, so too is the assembly of the ICL. As we will explore further in subsequent sections of this chapter, the assembly history of BCGs is closely intertwined with that of the ICL. This interconnectedness underscores the complex and dynamic nature of galaxy cluster environments, where the formation and evolution of both BCGs and the ICL are intricately linked.

\section{Intracluster light}\label{chapBCGICL:ICL}
The ICL, also known as diffuse light (DL), is a distinctive component of galaxy clusters and groups. Initially discovered by \cite{zwicky1937}, it has garnered significant attention in recent years from numerous authors (\citealt{zibetti2005,sommer-larsen2006,rudick2009,krick2007,puchwein2010,burke2012,martel2012,contini2014,demaio2015,iodice2017,groenewald2017,morishita2017,mihos2017,ko2018,spavone2018,tang2018,montes2018,jimenez-teja2019,zhang2019,demaio2020,yoo2022,ragusa2023,joo2023,contini2023,contreras-santos2024}, among others. For a more comprehensive list of references, please consult the reviews by \citealt{contini2021}, \citealt{arnaboldi2022} and \citealt{montes2022}).
It quickly became evident that the ICL is a crucial factor to consider when modeling the formation and evolution of galaxies using numerical simulations and/or semi-analytic models (e.g., \citealt{puchwein2010,contini2014}). This is particularly significant as the ICL can constitute up to 40\% or more of the total light within the virial radius of clusters. Beyond its role in contributing to the overall stellar budget, studying its properties can provide insights into the mechanisms driving the assembly of galaxy groups and clusters.

\begin{figure}[t]
\centering
\includegraphics[width=.7\textwidth]{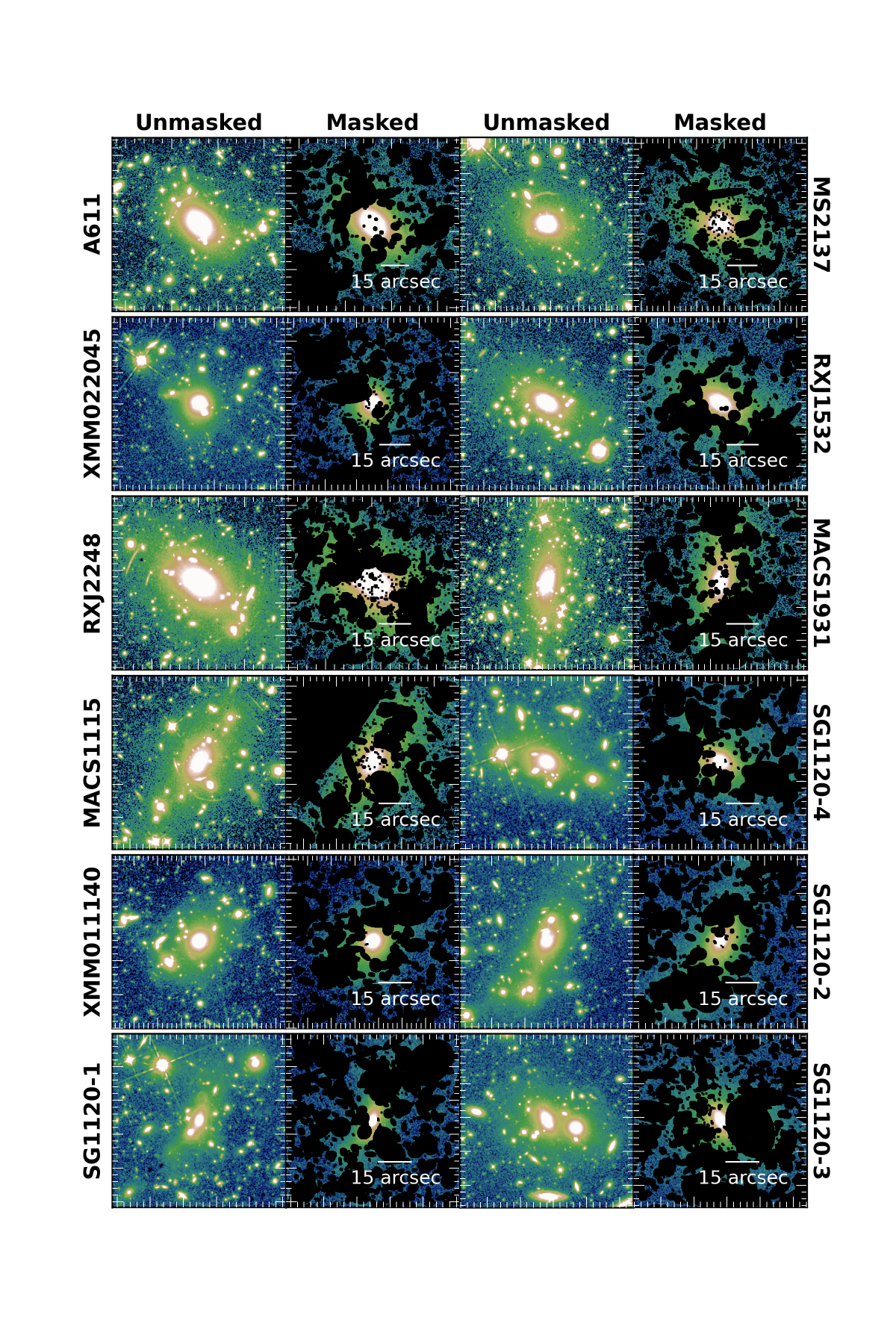}
\caption{The innermost 200 kpc of the masked and unmasked F160W images are depicted for 12 groups and clusters within the redshift range $0.29<z<0.89$. In the unmasked images, pixels brighter than 22 mag arcsec$^{-2}$ are represented in white, while black regions in the masked images indicate masked sources. Credit: \cite{demaio2018}.}
\label{chapBCGICL:ICLdistr}
\end{figure}

As mentioned previously, the ICL consists of stars that are not gravitationally bound to any specific galaxy within the cluster and are influenced solely by the potential well of the dark matter halo. These stars were originally part of satellite galaxies that either merged with the BCG, were stripped away while orbiting around it, or were accreted from other objects during the formation of the cluster (\citealt{ragusa2023,contini2024a}). Experts in this field suggest that the primary mechanisms responsible for the formation of the ICL are threefold: mergers between satellites and the BCG, the stripping of stars from satellites due to tidal forces during their orbits, and the accretion or pre-processing of material from external sources. Although \emph{in situ} star formation was initially proposed (\citealt{puchwein2010}), observational evidence later refuted this idea (\citealt{melnick2012}), indicating minimal ongoing star formation in the regions occupied by the ICL. Further elaboration on these processes will be provided in Section \ref{chapBCGICL:ICL_fm}.

The ICL can constitute a substantial fraction of mass or light within the virialized region of galaxy clusters, typically ranging from 5\% to 40\%, with no discernible trend in relation to cluster mass (\citealt{montes2022,ragusa2023}) or redshift (\citealt{joo2023,contini2024a}). Investigating the role of the ICL in determining the total stellar mass within clusters, particularly the scatter in this relationship, can provide insights into the dynamical state of the clusters under study (\citealt{ragusa2023,contini2023,jimenez-teja2024}). Our research has focused on studying various properties such as metallicity, colors, and age to gain a deeper understanding of the formation mechanisms of the ICL. For instance, if the properties of the ICL closely resemble those of the BCG, it suggests shared formation mechanisms. An illustrative example of this, which we will revisit in Section \ref{chapBCGICL:BCGICL_prop}, is the presence or absence of gradients in colors and/or metallicity across both the BCG and the ICL. The absence of such gradients may imply that mergers are the primary formation mechanism, while the presence of gradients could suggest a more gradual mechanism such as stellar stripping (\citealt{contini2019}).

An important challenge in studying the ICL arises from its diffuse nature, as it permeates regions that may also be occupied by stars still bound to other galaxies. Consequently, separating the ICL from other components presents a significant challenge. In numerical simulations or semi-analytic models, the distinction between ICL stars and others is clear, thanks to available dynamical and kinematic information. However, in observations, this task becomes daunting due to the lack of such detailed information. Separating the ICL from the rest, including the BCG, is particularly challenging, to the extent that many authors opt to study the entire BCG+ICL system without attempting separation (\citealt{contini2021,montes2022} and references therein). Despite these definitional challenges, the ICL serves as a valuable tool for understanding the formation and assembly of clusters by linking its properties with their dynamical stages. Below, we explore how recent studies utilize the ICL to investigate the dark matter content in clusters.

Before delving into the mechanisms responsible for the formation of the ICL, it is worthwhile to examine the typical distribution of the ICL in clusters. This is illustrated in Figure \ref{chapBCGICL:ICLdistr}, adapted from \citealt{demaio2018}, which depicts the innermost 200 kpc of 12 groups and clusters at the lowest redshifts in their sample. The authors differentiated the contribution of the ICL by masking satellite galaxies and provided both unmasked and masked images. A comparison between the two images for the same clusters reveals a significant remaining light, which includes the contribution from the BCG. In Section \ref{chapBCGICL:BCGICL_sep}, we will explore potential techniques for separating the ICL from the BCG.

\subsection{Formation mechanisms}\label{chapBCGICL:ICL_fm}
The growth of the ICL can be attributed to several channels, with stellar stripping and mergers being the primary processes investigated (\citealt{monaco2006,murante2007,somerville2008,guo2011,martel2012,contini2014,contini2018}). Stellar stripping occurs when stars belonging to a satellite galaxy are stripped away by tidal forces acting on the satellite itself, while mergers involve the unbinding of a fraction of the satellite's mass during the merger process with the BCG. However, stellar stripping and mergers are not the only mechanisms proposed. Additionally, other processes must be considered, including the disruption of dwarf galaxies (\citealt{conroy2007,purcell2007,giallongo2014,contini2014,annunziatella2016,raj2020}), \emph{in situ} star formation (\citealt{puchwein2010}), and the accretion of ICL already formed elsewhere (\citealt{contini2014,contini2023,joo2023,ragusa2023}), also known as pre-processing.

The disruption of dwarf galaxies is a common process occurring within groups and clusters. Physically, it is akin to stellar stripping, but in this scenario, the tidal forces are so intense that the galaxy is completely destroyed. This phenomenon primarily occurs near the cluster center where tidal forces are strongest, and at low redshifts when halos are generally more concentrated compared to their higher redshift counterparts (\citealt{gao2011,contini2012}). Despite the frequency of dwarf galaxy disruption, studies (\citealt{contini2014}) have demonstrated that it contributes minimally to the stellar mass gained by the ICL. Indeed, research by various authors (e.g., \citealt{montes2018,contini2018,contini2024a}) has shown that stripping of intermediate/massive galaxies ($\log M_* > 10.5$) accounts for approximately 70\% of the material originating from stellar stripping. Consequently, the disruption of dwarf galaxies has been discounted as a primary channel in the formation of the ICL.

\emph{In situ} star formation was initially observed in simulations by \cite{puchwein2010}, where it was found to contribute significantly (about 30\%). However, subsequent observational evidence (\citealt{melnick2012}) contradicted these findings, indicating that \emph{in situ} star formation contributes no more than approximately 1\%. It is worth noting that, at the time of writing this chapter, a study by \cite{ahvazi2024} claims that \emph{in situ} star formation in three massive clusters from the TNG50 simulations can account for approximately 8\% to 28\% of the total ICL. Furthermore, this study suggests that the majority of these stars are formed directly within the ICL, hundreds of kiloparsecs away from the center of the host halo. It should also be noted that there are some claims that stars can be formed in the intracluster medium, either through stripped gas, or directly in gas (e.g., \citealt{gullieuszik2020}).

Pre-processing or the accretion of already processed ICL is considered to be a significant channel for the formation of the ICL. Pre-processed ICL refers to the ICL formed within some groups that later merge during the assembly of a given cluster. During these mergers, the groups also bring along their own ICL contents, which merge with the cluster's ICL once the merger is complete. Several studies (e.g., \citealt{contini2024a} and references therein) have demonstrated that pre-processed ICL can contribute up to 30\%-40\% of the total ICL in clusters at the present time. However, it is important to note that pre-processing is not a direct means of forming ICL (\citealt{contini2024a}). The ICL primarily forms through the processes described earlier, and part of it can be accreted during the assembly history of clusters. Therefore, this channel is essentially a sub-channel (see, e.g., the discussion in \citealt{contini2024a}).

Stellar stripping and mergers are now recognized as the two most significant channels in the formation of the ICL, with pre-processing considered a sub-channel. However, quantifying the contributions of stellar stripping and mergers separately is challenging, particularly due to the dependency on the definition of a merger. Authors often have varying definitions of what constitutes a merger, especially concerning the timing of when a merger begins.
For instance, \cite{murante2007} asserted that approximately 75\% of the ICL forms through mergers, while \cite{contini2014} found that around $\sim 80$\% originates from stellar stripping. This discrepancy arises from the authors' adoption of different definitions of a \emph{merger}. If the timing of when a merger begins is not uniformly defined, the contribution from mergers can result in varying percentages. In \cite{contini2014}, and generally in semi-analytic models, a merger is defined as the moment when the satellite can no longer be distinguished from the central galaxy, effectively becoming the same object. However, while orbiting very close to the central galaxy, the satellite may undergo significant tidal stripping. In this scenario, the stripped stars can also be considered as originating from the merger channel if the satellite is on the verge of merging. In a subsequent study, \cite{contini2018} demonstrated that by relaxing their definition of a merger to align more closely with that of Murante et al., which was based on information in the merger trees, the percentage of ICL attributed to the merger channel converged with that claimed by Murante et al..

It is important to note, however, that since the work of Murante et al., numerical simulations have made significant advancements in terms of accuracy and mass resolution. It is pertinent to revisit this point with more resolved simulations available today. For instance, leveraging the TNG300 simulation of the IllustrisTNG project (see \citealt{pillepich2018} and references therein), \cite{montenegro-taborda2023} demonstrated that a fraction ranging from 50\% to 60\%, depending on the mass of the host halo, of the ICL originates from the merger channel, encompassing very minor, minor, and major mergers.

How can we quantify the respective contributions of mergers and stellar stripping to the formation of the ICL? One approach, commonly employed in observations, involves studying the main properties of the ICL itself, without the need for separation from the BCG. Mergers and stellar stripping are expected to imprint distinct characteristics on the ICL, thus examining its properties (or considering the entire BCG+ICL system) allows us to discern which process plays the most significant role. In the subsequent section, we will explore how certain properties of the ICL can provide insights into this matter.

\section{BCGs and ICL properties}\label{chapBCGICL:BCGICL}
The only viable approach, observationally, to gain insight into the mechanisms underlying the formation of BCGs and ICLs is by examining their properties. However, this poses significant challenges in observations, as isolating the BCG+ICL from the contributions of satellites is inherently difficult. Additionally, the ICL is a faint component that may even fall below the sky level, further complicating its isolation. Before delving into the most important (and studied) properties of BCGs and ICL, we will address in the next sub-section the primary techniques for separating the ICL from the BCG. This step is crucial for understanding the distribution of the two components and, more importantly, recognizing the absence of a specific observable quantity capable of delineating a clear boundary between them without some compromise.

\subsection{Separation of the two components}\label{chapBCGICL:BCGICL_sep}
The separation of the ICL from the BCG presents significant challenges in observational studies, whereas it is comparatively easier in theoretical approaches. Simulations and semi-analytic models have the advantage of utilizing dynamical and kinematic information, which greatly facilitates the task when coupled with a reliable definition of the ICL. For instance, in simulations, we possess detailed information regarding the positions and velocities of every individual star particle, allowing for the straightforward separation of the ICL from the BCG via the velocity distribution of the BCG+ICL stars (see, e.g., \citealt{dolag2010}). Using this method, it becomes evident that the total velocity distribution is not simply governed by a single Maxwellian distribution, but rather a combination of two Maxwellian distributions—one describing the distribution of stars belonging to the BCG and the other characterizing the stars belonging to the ICL.

However, the velocity distribution method is generally not applicable in observational studies due to the lack of information on individual stars. A similar approach, albeit limited to nearby objects, involves examining the bimodality of the velocity distribution of Planetary Nebulae (\citealt{longobardi2015}). Their velocity distribution can be segregated into a narrow component associated with the BCG and a broader one linked to the ICL, thereby allowing for an estimation of the transition region between the two components. Unfortunately, this method is often impractical due to the distances of most observed targets from us.

Among the observational methods commonly used, two noteworthy techniques are the isophotal limit cut-off (e.g., \citealt{zibetti2005}) and profile fitting methods (e.g., \citealt{zhang2019}). These methods differ fundamentally, often resulting in different separations when applied to the same target. Both techniques rely on thorough masking of external sources. The isophotal limit cut-off method involves a \emph{subjective} cutoff in surface brightness, below which all remaining light is considered ICL. This simplistic approach does not account for the ICL in the transition region between the two components and is susceptible to contamination from massive galaxies within the clusters.

Another widely used method in observations is profile fitting, which employs functional forms to fit the BCG+ICL light. This method typically assumes at least two components to describe both the BCG and ICL, but a minimum of three is necessary to characterize the transition region between them. The triple Sersic profile fitting is commonly used for this purpose. While profile fitting offers the advantage of describing the transition between the BCG and the ICL, it is heavily reliant on the specific functional forms chosen.

It is worth noting that recent studies, both observational and theoretical, have adopted a trend of not attempting to separate the two components (e.g., \citealt{montes2018,pillepich2018,contreras-santos2024}). In these cases, a simple assumption is made to consider a certain distance from the BCG where the light is assumed to originate from the BCG, and beyond this distance, from the ICL. While this approach completely overlooks the transition region between the two components, it can be considered reasonable if the assumed distance lies within this region. Another reliable criterion for determining this distance is the transition radius between the two components, as suggested by \cite{contini2022} (but see also \citealt{proctor2024,brough2024}). This radius can be quantified in observations via profile fitting and easily derived in theoretical models.

\subsection{Fraction of ICL in galaxy clusters}
The quantification of ICL mass is perhaps one of the most straightforward properties to assess. Assuming a reasonably trustworthy separation of the ICL from the BCG, the significance of the ICL in galaxy groups and clusters can be evaluated by plotting the ICL mass over the total stellar mass within a specified radius from the cluster center, as a function of cluster mass within a given overdensity. Alternatively, to avoid the separation from the BCG, several studies have considered the stellar mass in BCG+ICL over the total (e.g., \citealt{gonzalez2013}). By compiling the observed fraction of ICL mass, it becomes apparent that the ICL cannot be disregarded. Indeed, it can account for as much as 40\%-50\% of the total stellar mass within the virial radius of a group or cluster (see the review by \citealt{montes2022}).

There is no general consensus on whether the ICL fraction depends on the halo mass, but most studies to date suggest that it is independent of halo mass, at least for $\log M_{\rm{halo}} \geq 13$ (see \citealt{contini2024b} for results in a lower halo mass range). However, the relation is characterized by a significant scatter (e.g., \citealt{contini2021}). If this is indeed the case, it implies that the physical processes discussed above, responsible for ICL formation, are similarly efficient regardless of the mass of the host. Furthermore, given the observed scatter in the relation, which is also predicted by several theoretical models (e.g., \citealt{murante2007,guo2011,contini2014,pillepich2018,contini2023}), it is essential to understand which halo properties can account for it. We will revisit this point shortly. An important caveat to consider when comparing the ICL fraction derived from different studies, whether observational or theoretical, is the radius within which the ICL fraction is computed. While most ICL fractions are computed within the virial radius of the halo, and attention must be paid to the definition of the virial radius, others are computed within a specified distance from the center, especially in observational studies constrained by the limits of the observation (e.g., \citealt{montes2018}).

\begin{figure}[t]
\centering
\includegraphics[width=.7\textwidth]{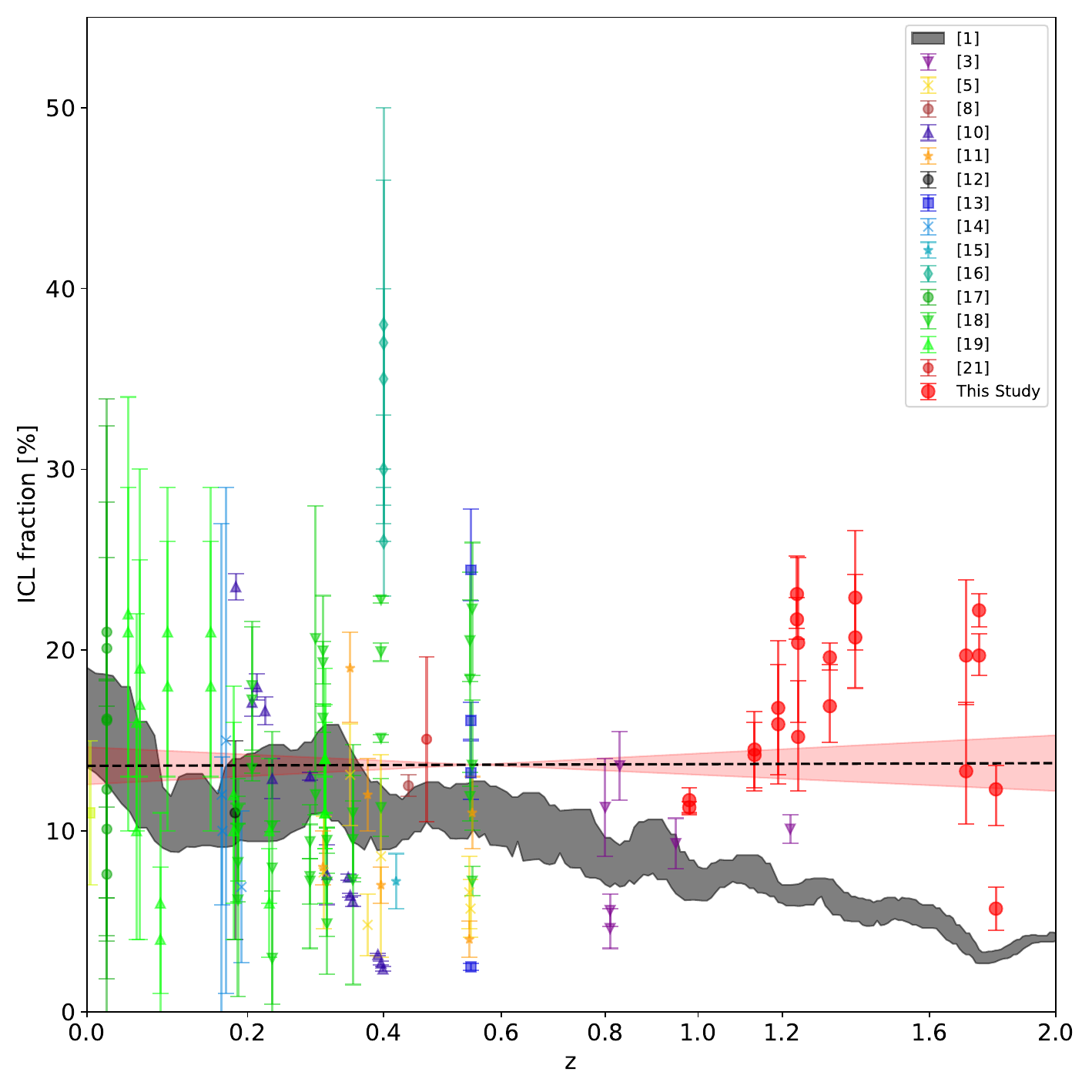}
\caption{A collection of ICL fractions at different redshifts, ranging from the present day to redshift $z\sim 2$, is presented. The plot clearly illustrates that the ICL fraction is already significant at high redshifts and comparable to that observed in the local Universe. References: (1) \cite{rudick2011}; (3) \cite{burke2012}; (5) \cite{montes2018}; (8) \cite{presotto2014}; (10) \cite{burke2015}; (11) \cite{morishita2017}; (12) \cite{alamo2017}; (13) \cite{ellien2019}; (14) \cite{feldmeier2004}; (15) \cite{griffiths2018}; (16) \cite{jee2010}; (17) \cite{jimenez-teja2018}; (18) \cite{jimenez-teja2019}; (19) \cite{krick2007}; (21) \cite{yoo2021}, and they point labelled as \emph{This study} are from \cite{joo2023}. Credit: \cite{joo2023}.}
\label{chapBCGICL:iclfrac_red}
\end{figure}

A crucial aspect related to the formation of the ICL pertains to its evolution over time in terms of the ICL fraction. Theoretical studies generally concur (e.g., \citealt{murante2007,contini2014,contini2024a}) that the ICL forms relatively late. However, there is mounting observational evidence indicating that the ICL fraction is already substantial at high redshift (\citealt{ko2018,joo2023}). These two points are not necessarily contradictory, given that (a) the term 'formation time' typically refers to when the bulk of the ICL is established and (b) at high redshifts, while the amount of ICL may be lower, the total stellar mass within a halo is also lower compared to more evolved objects in the present universe. Figure \ref{chapBCGICL:iclfrac_red}, extracted from \cite{joo2023}, presents various observed ICL fractions at different redshifts. The figure illustrates that high-redshift objects harbor comparable amounts of ICL relative to lower-redshift counterparts, implying that the ICL is already prominent at high redshifts (relative to the total stellar mass). Recent theoretical studies, such as \cite{contini2024a}, have demonstrated that models can reproduce the observed fractions and their redshift dependence, aligning with the principal findings of \cite{joo2023}. Indeed, while previous theoretical studies were focused on the assembly history of the ICL, which means selecting a given object at the present time and read its history through the merger tree in order to derive the ICL formation time, more recent theoretical studies such that quoted above selected massive groups and clusters directly at high redshifts. The two selections are very different in their essence.

\begin{figure}[t]
\centering
\includegraphics[width=.7\textwidth]{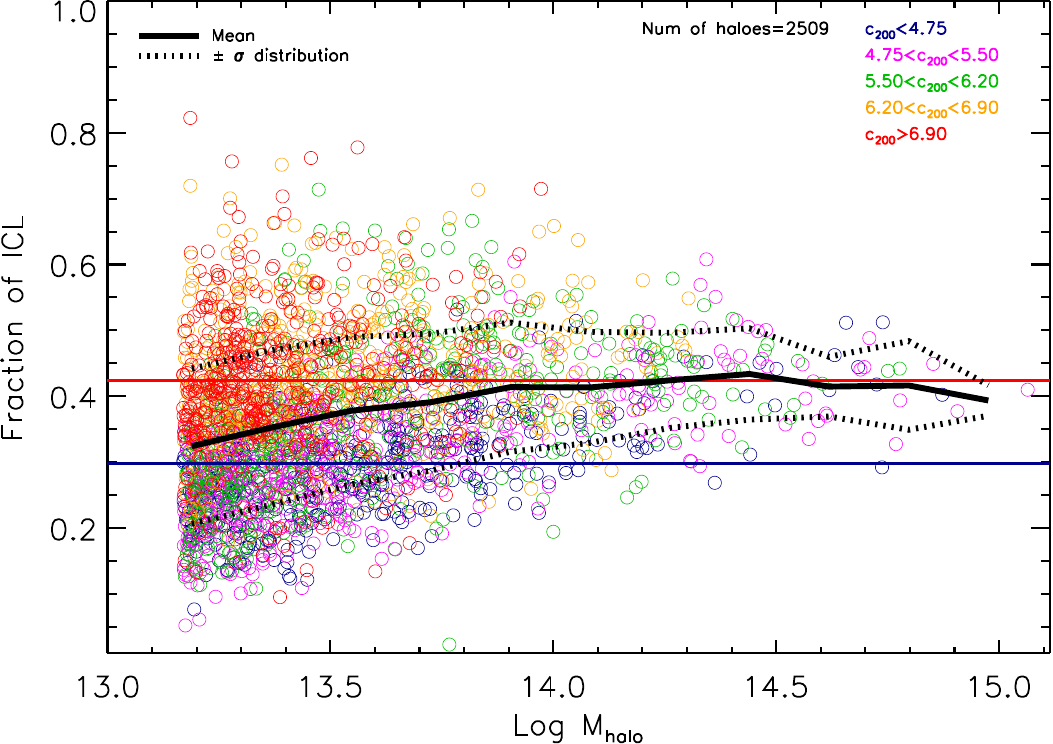}
\caption{The fraction of ICL as a function of halo mass is depicted, with color-coding based on the concentration of the halo. At a given halo mass, more concentrated halos tend to host a higher fraction of ICL. The red and blue lines represent the average fraction of the halos in the highest and lowest concentration ranges, respectively. Credit: \cite{contini2023}.}
\label{chapBCGICL:iclfrac_conc}
\end{figure}

The scatter in the ICL fraction-halo mass relation is crucial for explaining the primary driver of ICL formation, yet it remains a topic of debate. At a given halo mass, the ICL fraction can vary from a few percent to roughly 50\%, indicating significant variability within halos of similar mass. How can we account for this variability? It likely depends on factors related to either the physical mechanisms driving ICL formation or certain halo properties, possibly linked to their assembly history. As discussed earlier, mergers and stellar stripping are the primary channels for ICL formation. Regardless of which process dominates, they are expected to operate differently in halos of similar mass but with different ICL content. One possible explanation is that halos in a more evolved dynamical stage tend to have a higher fraction of ICL (Ragusa et al., 2023; Contini et al., 2023). This is because more evolved halos formed earlier, providing more time for ICL formation. In Figure \ref{chapBCGICL:iclfrac_conc}, adapted from Contini et al. (2023), the ICL fraction is plotted against halo mass, color-coded based on halo concentration. Halo concentration is defined as the ratio of the virial radius to the scale radius assuming an NFW profile (Navarro et al., 1997). Notably, the plot reveals significant scatter, particularly on group scales (larger scales are limited by statistics). The scatter predicted by their model aligns with observed trends across a wide range of halo masses. Importantly, the color-coding indicates that, on average, more concentrated (redder) halos exhibit a higher ICL fraction. This suggests that more dynamically evolved halos tend to have a higher proportion of ICL than less concentrated counterparts.

How can we explain the connection between the dynamical state of the halo and its ICL fraction within the context of ICL formation? The answer lies in the role of stellar stripping and mergers in ICL formation. If stellar stripping, as predicted by the model of Contini et al. (2023) and supported by many observations, is the primary mechanism, then the stronger tidal forces in more concentrated halos would lead to a higher ICL fraction. Additionally, massive satellites experience faster dynamical friction, allowing them to reach the most concentrated regions earlier than less massive satellites. Consequently, they are more susceptible to stellar stripping and contribute more to the ICL content. In the following subsection, we will explore how observational results on the main properties of ICL and BCGs shed light on the dominant mechanisms and the types of satellites involved.

\subsection{Colors, metallicity and age}\label{chapBCGICL:BCGICL_prop}
The formation and evolution of BCGs and ICL are interconnected, as the ICL is often regarded as an envelope surrounding the BCG with which it is associated. However, since the ICL consists of stars not bound to the BCG itself, their formation histories may differ, potentially experiencing growth in stellar mass at different times, driven by distinct mechanisms. We will delve into this point shortly, but it is crucial to clarify that observationally, the only way to understand the role of these mechanisms in the growth of both components is by examining their properties, particularly their colors, metallicities, and ages. Despite the complexities arising from the age-metallicity degeneracy, these parameters are essential for deciphering which mechanisms predominantly contribute to the growth of the ICL.

Having outlined the two primary mechanisms for ICL formation, alongside the significant mode of mass growth through pre-processing, we can now explore how mergers and stellar stripping fit into various scenarios where colors, age, and metallicity may be similar or different in BCGs and ICL. Let's focus on three possible scenarios that emerge when measuring one of these properties in regions dominated by the BCG compared to those where the ICL dominates. Notably, we can analyze these scenarios without needing to separate the two components, thus avoiding several complications associated with such separation. For simplicity, we assume that the BCG dominates the central regions, while the ICL becomes prominent at greater distances from the center. Let's consider the distribution of colors (although the same logic applies to the other properties) in the BCG+ICL as a whole, as a function of distance from the central regions. We can envision three possible scenarios:

\begin{itemize}
\item [(1)] colors become bluer with increasing distance;
\item [(2)] colors become redder with increasing distance;
\item [(3)] colors remain constant regardless of the distance from the central regions.
\end{itemize}

In the first scenario, we observe a net negative gradient from the center to larger distances, while in the second scenario, the gradient is positive. In the last case, there is no change in colors, resulting in no gradient at all. Each of these possibilities offers insights into how mergers or stripping might influence the observed color distribution.

Violent relaxation processes during mergers are quite strong (\citealt{murante2007}), sufficiently robust to mix up the stellar populations. When a satellite merges with the BCG, some stars originally belonging to the satellite become unbound and join the ICL component. Depending on the magnitude of the merger (minor or major), the BCG's disk may be completely destroyed. In any case, mergers have the power to mix up stellar populations, making the last scenario the expected color distribution following a recent merger. Several observations support this idea (\citealt{raj2020,ragusa2021,joo2023}). Conversely, several other studies (\citealt{demaio2015,iodice2017,morishita2017,montes2018}) support a gradient in the color distribution.

In the case we observe a gradient in the color distribution, stellar stripping is the first candidate at play. Indeed, contrary to mergers, this mechanism is more gentle, as it occurs when the satellite and BCG are not physically connected, and only part of the satellite's stars are effectively stripped, mostly the outermost ones. The gradient, whether positive or negative, depends on the type of satellites undergoing stripping. Bluer colors indicate relatively younger stars, suggesting that, overall, the ICL stellar population is younger than that of the BCG, with these stars mainly originating from the outskirts of intermediate or massive satellites.

In present-day groups and clusters, the typical observation reveals a negative gradient in colors or metallicity (as indicated in the references above), i.e., a bluer and more metal-poor ICL. However, there are instances of flat distributions, which may result from recent major mergers or multiple minor mergers (e.g., \citealt{contini2019,joo2023}). Nonetheless, while stellar stripping may explain most observations for nearby objects, the scenario can differ at higher redshifts. With mergers being more frequent, the likelihood of a mix-up of the stellar population increases. This was demonstrated in a recent study of the ICL in objects at high redshift by \cite{joo2023}, who investigated ten galaxy clusters in the redshift range $1<z<2$ using deep IR imaging data. They found that, consistent with present-day observations, the ICL fraction in these clusters is comparable to that found in the present Universe. Moreover, they observed no clear correlation between colors and cluster-centric distance, leading them to conclude that stellar stripping is not the main mechanism at play in these high-redshift environments.

It is worth noting that, although most cases with flat color/metallicity distributions can be explained as results of mergers, in particular situations, stellar stripping can also be invoked. If the typical colors of the largest contributors to the ICL in a given cluster are similar to those of the BCG, stellar stripping of these satellites would inevitably result in a flat color distribution. This implies that flat color/metallicity distributions are not always evidence of recent mergers.

\section{ICL distribution in groups and clusters}\label{chapBCGICL:ICL_distr}
The measurable properties of the ICL serve as valuable tools for understanding not only how it was formed and assembled over time through the processes described above, but also as indicators of its distribution within dark matter halos. This characteristic is pivotal in ICL studies because, comprising stars not gravitationally bound to any galaxy and influenced solely by the potential well of the dark matter halo, it provides insights into the properties of dark matter. Specifically, the ICL can serve as a luminous tracer of the dark matter distribution (\citealt{yoo2024} and references therein).

The concept of linking the distribution of the ICL with that of dark matter is relatively recent (\citealt{montes2019}), although initial attempts were made several years ago. For instance, \cite{zibetti2005} were among the first to propose that the ICL distribution could be approximated by an NFW profile, while \cite{jee2010} utilized the ICL along with weak lensing to trace dark matter structures in clusters. Over time, various authors have explored connections between the ICL and dark matter halos (\citealt{contini2021}), but only in recent years has this approach been developed with a high level of precision and examined from different perspectives.

Fundamentally, these studies operate under the implicit assumption that since dark matter particles are collisionless and the ICL is bound only to the halo's gravitational potential, the distribution of ICL stars may resemble that of dark matter. This assumption suggests that the two components could be described by similar profiles, as assumed in recent theoretical investigations (\citealt{contini2020,contini2021}). Initially, \cite{harris2017} concluded that the ICL was more centrally concentrated than dark matter. This finding was subsequently corroborated by studies such as those by \cite{contini2020} and \cite{contini2021}, which assumed an NFW profile for the ICL with a higher concentration compared to the dark matter halo. These studies demonstrated that by assuming a more concentrated ICL distribution, their models could reproduce the observed BCG+ICL mass within various apertures and across different redshifts. Moreover, their model was used to predict the typical radius of transition between the BCG and ICL (\citealt{contini2022}), providing a method for separating these two components

The most significant findings, however, arise from observational studies (though supported by theoretical evidence). For instance, \cite{montes2019} quantified the average separation between the ICL and dark matter distributions in six clusters from the Hubble Frontiers Field. Utilizing the Modified Hausdorff distance, which gauges the average distance between two distributions, they determined that the mean separation between the total mass distribution and the ICL is approximately $\sim 25$ kpc. Furthermore, the ICL proved to be more adept at describing the dark matter distribution than previous proxies, such as X-ray measurements. Subsequently, \cite{asensio2020} and \cite{yoo2022} theoretically corroborated these results using similar techniques.

Numerous recent studies have endeavored to establish a link between the ICL and dark matter (\citealt{pillepich2018,zhang2019,deason2021,sampaio2021,kluge2021,chun2023,contini2023,contini2024a,jimenez-teja2024}). To summarize a few, \cite{zhang2019} investigated a cluster sample from the Dark Energy Survey and identified a universal radial dependence of the BCG+ICL surface brightness. However, they concluded that there was insufficient evidence to support the ICL as a reliable tracer of the cluster matter distribution, but rather as an excellent indicator of the total cluster mass. Meanwhile, \cite{kluge2021} focused on the alignments of the BCG and ICL with their host dark matter halo, finding that the ICL aligns more closely than the BCG, thus proposing the ICL as a dependable tracer of the dark matter distribution (see also \citealt{ragone2020}).

Recent observational and theoretical works have also explored the relationship between the ICL and the dynamical state of the host halo (\citealt{ragusa2023,contini2023,contini2024a,jimenez-teja2024,yoo2024}). The underlying concept is that if the ICL forms through mechanisms associated with the overall assembly of the halo, its properties should be linked to the particular growth stage and/or dynamical state of the halo. For instance, \cite{jimenez-teja2024} observed a higher ICL fraction in clusters that experienced a major merger relatively recently compared to those in a more advanced dynamical stage. This finding contrasts with the results of another observational study by \cite{ragusa2023}, where a higher ICL fraction was observed in groups and clusters with a greater proportion of early-type galaxies, typically found in more evolved objects. These last results align with the theoretical predictions of \cite{contini2023,contini2024a}, who found a higher average ICL fraction in halos that are more concentrated and formed earlier than others, a trend later extended to Milky Way-like halos (\citealt{contini2024b}).

\section{Evolution with time of BCGs and ICL}\label{chapBCGICL:BCGICL_coev}
BCGs and ICL, despite being distinct entities, share common pathways of formation. It has been demonstrated (e.g., \citealt{contini2018}) that at least part of their formation is attributed to physical mechanisms contributing to both their growth, namely stellar stripping and mergers. The extent to which each mechanism contributes to the two components remains a topic of debate, not only between observers and theorists but also among authors within the same field. It is reasonably certain that BCGs undergo at least two growth phases: a rapid one at high redshifts due to mergers, and a smoother accretion at lower redshifts. Here we are excluding the initial stage of intense star formation (which is distinct from ICL formation) happening at high redshifts (before $z\sim 2-3$), depending on the mass of the galaxy (see, e.g., \citealt{delucia2007}). Most theories predict a late formation of the ICL, implying that mechanisms driving early BCG growth may not fully account for the initiation of ICL formation. However, it is evident that regardless of the primary mechanism driving ICL formation, it must also play some role in BCG growth. For instance, if mergers or smooth accretion primarily drive ICL growth, they also contribute to BCG growth at later times. Conversely, if stellar stripping is predominant, galaxies subject to stripping are destined to merge with the BCG. In essence, while BCGs and ICL are distinct components, they share common features in their formation, as discussed below.

Beginning with the evidence that the ICL is dynamically bound to its host (\citealt{montes2018,yoo2022}) and the understanding that more massive halos host the most massive BCGs (e.g., \citealt{moster2018}), we can infer that the growth of both BCGs and ICL are somehow linked to the dynamical evolution of the host halo. Under this premise, BCGs and ICL should evolve similarly from a certain point in time, which is pivotal for the ensuing discussion. The hierarchical manner in which BCGs accumulate mass entails that they acquire most of their stellar mass at high redshifts, with their growth diminishing rapidly after redshift $z\sim 1$ (e.g., \citealt{burke2015}). Coinciding with this, numerical simulations and semi-analytic models (e.g., \citealt{murante2007,contini2014}) suggest that around redshift $z\sim 1$, the ICL undergoes an acceleration in its growth. To quantify the growth of BCGs and ICL after this redshift, which roughly corresponds to the last 7-8 billion years, numerical methods indicate that the ICL accumulates approximately 80\% of its stellar mass by the present time. In contrast, observations yield varying estimates for BCG growth during this period, with some claiming no growth (\citealt{oliva-altamirano2014}), approximately 50\% growth (\citealt{lin2013,burke2015}), and others suggesting a doubling in mass (\citealt{lidman2012}). However, it must be noted that the selection of the progenitors in observations is not trivial, and might be affected by progenitor bias, i.e., the assumption that high redshift galaxies are drawn from the same distribution as the low redshift galaxies.

\begin{figure}[t]
\centering
\includegraphics[width=.7\textwidth]{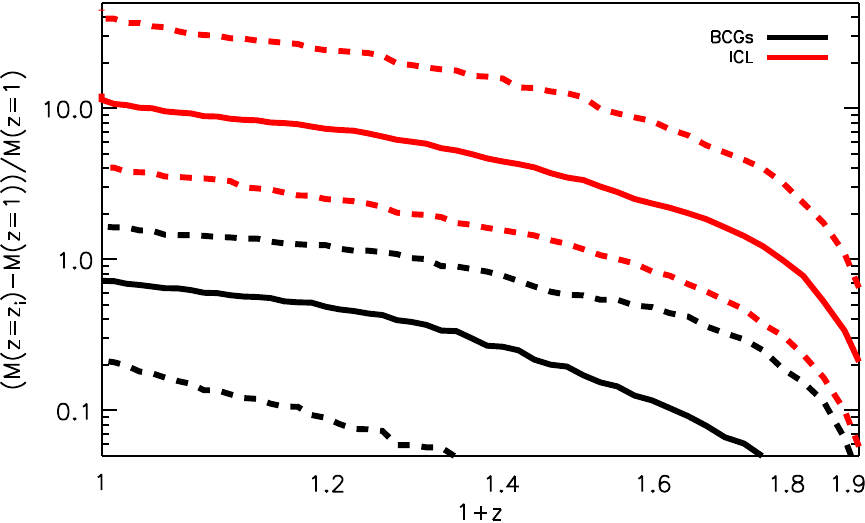}
\caption{The residual mass of the ICL (red lines) and BCGs (black lines) normalized by the mass at redshift $z=1$. Solid lines represent the median, while the dashed lines represent the 16th and 84th percentiles of the distributions. Credit: \cite{contini2018}.}
\label{chapBCGICL:iclbcg_growth}
\end{figure}

The evolution of BCGs and ICL over the last 7-8 billion years has been extensively examined by \cite{contini2018} using their semi-analytic model, which enables the tracking of the growth of both components and elucidates whether they co-evolve. Co-evolution in this context does not necessarily imply that BCGs and ICL assemble the same amount of stellar mass during this period, but rather that they do so at similar rates, irrespective of the total mass accumulated, and through the same underlying mechanisms. Figure \ref{chapBCGICL:iclbcg_growth} from \cite{contini2018} illustrates the residual mass of the ICL (red lines) and BCGs (black lines) normalized by their masses at redshift $z=1$. Solid lines represent the median, while dashed lines indicate the 16th and 84th percentiles of the distributions. Normalizing at $z=1$ ensures that the earlier growth of the two components is cancelled out, allowing for a focus on their differential growth rates. Essentially, the figure quantifies the rate at which BCGs and ICL grow, with the slope of the lines at each redshift indicating the acceleration of this growth rate. The authors demonstrated that the slopes of the black (BCGs) and red (ICL) lines become equivalent after $z\sim 0.7$, indicating that BCGs and ICL co-evolve beyond this redshift.

The co-evolution of BCGs and ICL over the last few billion years has significant implications, aligning well with most of the observational findings mentioned previously. This co-evolution suggests that despite being physically distinct components, they share common formation pathways. The prevailing scenario posits that BCGs predominantly form stars through early star formation, followed by rapid growth via mergers. Concurrently, as halos become increasingly centrally concentrated over time, stellar stripping becomes more potent, initiating the formation of the ICL. As time progresses, BCGs acquire less mass compared to earlier epochs, while the ICL undergoes rapid growth. Nevertheless, mergers and stripping continue to contribute to both components, with the growth of the ICL accelerating in the last few billion years as dark matter halos become more concentrated or dynamically evolved. Consequently, a portion of the stellar mass in satellites destined to merge is subjected to stripping and eventually incorporated into the ICL.

\section{Concluding remarks}\label{chapBCGICL:conclusions}
The most relevant takeaway is that, despite being separate entities, BCGs and ICL share many commonalities, particularly in their growth over the last few billion years and, in some cases, similar properties such as colors, metallicity, and age. As discussed earlier, understanding their properties is crucial for unraveling their formation and evolution. Violent processes like mergers would homogenize the stellar populations of both components, resulting in similar colors or metallicities, whereas gentler mechanisms like stellar stripping of satellite galaxies would lead to gradients in these observables depending on the typical satellites that have been stripped. Despite some exceptional cases, the ICL is usually bluer, less metal-rich, and younger than the associated BCG.

Regarding BCGs, they represent the most massive galaxies in the Universe, and due to their unique location in galaxy groups and clusters, their growth differs significantly from the rest of the galaxy population. A rapid growth at high redshift due to intense star formation activity is followed by growth through mergers at lower redshifts. The studies mentioned above have highlighted that BCG growth can also be almost negligible in some cases after $z\sim 1$, but overall they grow by a factor of approximately 1.5-2 in the latter half of the Universe's lifespan. In contrast, the ICL has a different early history. While BCGs grow via mergers, the ICL begins to form through both mergers and stellar stripping. However, the bulk of the ICL forms after $z\sim 1$ on average because stellar stripping becomes more powerful over time, particularly as the concentration of the host halo increases. Haloes of the same mass at lower redshifts are more concentrated and dynamically evolved, leading to differences in the production of stray stars from high redshift to more recent times.

As mentioned earlier in the chapter, given the rapid advancement of numerical simulations in terms of accuracy and resolution, it is crucial to reassess the primary contributors to the ICL. While its significance depends on the definition of mergers, conducting a comprehensive study considering various definitions is feasible. We aim to conduct such a study ourselves, utilizing suitable simulations such as YZiCS (\citealt{choi2017}) or the upcoming simulation New Cluster (Han et al., in preparation). These simulations offer high resolutions, particularly New Cluster, making them suitable for investigating this topic.

Another crucial point of this chapter is the link between the ICL, or BCG+ICL content, and the dynamical state of their host. These components can serve as luminous tracers of the dark matter distribution, particularly the ICL, considering it is a collisionless component like dark matter. This approach holds promise for studying dark matter, understanding the intrinsic differences between relaxed and unrelaxed halos, as well as their assembly history. Recently, \cite{yoo2024} focused on the spatial distribution between dark matter and baryonic components such as gas, satellites, BCG+ICL, and stellar particles alone in a theoretical manner. Their study echoes previous works such as \cite{montes2019}, \cite{asensio2020}, \cite{yoo2022}, and others. By investigating the potential similarities in the distribution of dark matter and the mentioned components, they conclude that the spatial distribution of BCG+ICL and gas exhibits high similarity with that of dark matter, particularly for relaxed clusters (i.e., more evolved). This result implies once again that the ICL, together with the BCG, can be used to trace dark matter, serving as a probe of the dynamical state of the host halo.

\begin{definition}\label{def:mergers}
Mergers between dark matter halos are typically defined as the moment when the smaller halo enters the virial radius of the larger one, effectively becoming a subhalo (\citealt{springel2005}). The definition of a merger between galaxies, however, is somewhat controversial and lacks a universal standard due to the difficulty in determining the exact start and end times of the process. Consequently, variations exist across studies on a wide range of topics, including the formation of BCGs and the ICL. For simplicity, unless otherwise specified, in this chapter, we consider the merger concluded when the two galaxies are no longer distinguishable entities and merge into a single object.
\end{definition}

\begin{ack}[Acknowledgments]

E.C. and S.K.Y. acknowledge support from the Korean National Research Foundation (2020R1A2C3003769). E.C. and S.J. acknowledge support from the Korean National Research Foundation (RS-2023-00241934). All the authors are supported by the Korean National Research Foundation (2022R1A6A1A03053472).
\end{ack}


\bibliographystyle{Harvard}
\bibliography{reference}

\end{document}